# Deterministic Task Offloading and Resource Allocation in the IoT-Edge-Cloud Continuum

Keyvan Aghababaiyan, Baldomero Coll-Perales, Javier Gozalvez
Uwicore laboratory, Universidad Miguel Hernandez de Elche, Elche (Alicante), Spain
kaghababaiyan@umh.es, bcoll@umh.es, j.gozalvez@umh.es

*Abstract*— Future cellular networks will sustainably integrate computing, intelligence and services within a "network of networks" ecosystem that includes IoT devices and subnetworks for local communications and distributed processing. This integration creates an IoT-edge-cloud continuum that enables opportunistic task offloading across the continuum, enhancing network performance, reducing response times and allowing a flexible resource allocation that can facilitate the system to scale according to demand. Future networks should also natively support deterministic service levels for critical and time-sensitive vertical applications. In this paper, we propose a deterministic task offloading and resource allocation scheme for the joint management of communication and computing resources in the IoT-edge-cloud continuum. The proposed scheme prioritizes task completion before deadlines over minimizing the latency in the execution of individual tasks. The scheme leverages flexible latencies across tasks to support a higher number of tasks through a more efficient management of computing and communication resources that better adapts to scenarios with constrained resources.

## I. INTRODUCTION

Future 5G-Advanced and 6G networks are envisioned as a "network of networks" integrating diverse communication networks to provide seamless and ubiquitous connectivity [1]. This vision extends beyond traditional communication systems, aiming to seamlessly interconnect intelligence, computing and communication domains across an IoT-edge-cloud continuum. The IoT-edge-cloud continuum can offer access to additional and more powerful computational resources to meet the growing processing demands of deterministic and time-sensitive applications, which are shaping the communication and computing needs of future cellular networks. Many applications in critical (IoT) verticals, such as autonomous mobility and industrial automation, require deterministic processing capabilities to ensure adequate decision-making within bounded latency deadlines. These deadlines may range from stringent low-latency requirements to more relaxed limits that still need to be guaranteed. Providing deterministic end-to-end service levels is one of the key objectives in the design of future cellular networks [2]. The envisioned IoT-edge-cloud continuum will facilitate the seamless distribution of tasks across the continuum. However, it is critical to jointly manage communication and computing resources to ensure that neither a communication link nor a computing unit become overloaded, which would negatively impact system performance. This is particularly important for deterministic applications with bounded latency deadlines as effective offloading strategies in the continuum must consider the communication and processing demands of tasks, the availability of communication and computational resources across the continuum, and the processing and data transmission times.

To date, most task offloading and resource allocation proposals for the joint management of communication and computing resources have focused on minimizing the latency of individual tasks ([3]-[5]). While this approach might reduce latency for many tasks, it risks overloading certain computing and communication resources unnecessarily, potentially degrading overall system operation and negatively impacting the performance of other tasks. In contrast, this paper advocates for a deterministic approach to task offloading and resource allocation, prioritizing the completion of tasks within bounded deadlines (i.e., deterministic) over minimizing the latency of individual tasks. The proposal leverages the diverse latency requirements of deterministic tasks to efficiently support the maximum number of tasks through the joint management of communication and computing resources across the IoT-edge-cloud continuum. The paper presents the framework for the design of the deterministic task offloading and resource allocation, and formulates the proposal as an optimization problem constrained by the availability of communication and computing resources in the IoT-edge-cloud continuum. The study operates within the "network of networks" framework, integrating subnetworks for local IoT connectivity. The performance of the proposed deterministic solution is evaluated under various scenarios, considering different task loads and communication link qualities. The results demonstrate that a deterministic design for task offloading and resource allocation consistently supports a higher ratio of tasks across all evaluated scenarios compared to a state-of-the-art benchmark solution (following [4]) that focuses on minimizing latency.

## II. STATE OF THE ART

The integration of computing and intelligence enables an IoT-edge-cloud continuum, which provides opportunities to balance task allocation between local nodes and remote servers at the edge or in the cloud. While offloading tasks to remote servers can reduce computing or processing latency, it introduces communication latency due to data transmission from local nodes to remote servers. On the other hand, relying on local or IoT nodes may increase computing latency because of their generally lower processing power compared to edge nodes or cloud servers. It is important to note, however, that IoT nodes are not always resource-constrained devices. For instance, IoT devices such as Connected and Automated Vehicles (CAVs) or mobile robots often embed powerful

This work has been partially funded by the European Commission Horizon Europe SNS JU 6G-SHINE (GA 101095738) project, and by MCIN/AEI/10.13039/501100011033 (PID2020-115576RB-I00, PID2023-150308OB-I00).

processors, enabling a more balanced trade-off between executing tasks locally and offloading them to remote servers.

The opportunities provided by the IoT-edge-cloud continuum have spurred significant research in recent years to design task offloading and resource allocation mechanisms that optimize latency in distributed computing environments. Most existing contributions focus on minimizing total latency ([3]-[5]), which includes both computing and communication latency, by finding an optimal balance between local computation and task offloading to remote servers. Studies by Cai et al. [3] and Oliveira et al. [6] have explored this balance, highlighting that decisions regarding how much data should be processed locally versus offloaded remotely depend on factors such as communication bandwidth and available processing power. For instance, Cai et al. [3] demonstrated that optimizing computing and communication capacities in industrial IoT systems can significantly reduce latency through intelligent offloading strategies. Similarly, [7] proposed a task offloading framework using federated Q-learning to minimize both communication and computing latency. In [8], a reinforcement learning-based joint communication-and-computation resource allocation mechanism was introduced to optimize processing latency by jointly optimizing task offloading and resource allocation through a Q-learning-based online offloading algorithm and a Lagrange-based migration algorithm. In [9], the authors presented a multi-agent deep reinforcement learning (MADRL)-based computation offloading method capable of simultaneously addressing heterogeneous industrial IoT system requirements, including latency reduction. Oliveira et al. [6] proposed a task allocation strategy to minimize response times for latency-sensitive applications while reducing network traffic by mitigating idle resource time in hierarchical fog architectures supporting the IoT-edge-cloud continuum.

The advent of connected and autonomous mobility represents both an opportunity and a challenge for the IoT-edge-cloud continuum. On one hand, CAVs generate significant amounts of data that require timely processing. CAVs can exploit the continuum to offload non-critical and processing-intensive tasks to the edge or cloud, provided the offloading process does not compromise the communication latency. On the other hand, CAVs integrate powerful processing platforms, enabling a more balanced local-edge-cloud offloading of tasks, as local or IoT nodes can also support the processing of computationally-intensive tasks. Such balanced offloading is explored in [10], where the authors study a crowd sensing use case that opportunistically offloads data to the edge and analyzes various reverse offloading strategies to exploit vehicular and edge resources to reduce system latency. In [4], the authors propose a joint task offloading and resource allocation scheme that opportunistically offloads tasks to vehicles with unused computing resources, alleviating the load on edge servers and minimizing priority-weighted task processing latencies. A similar scenario is envisioned in [5] where the authors present a scheme that processes tasks locally or offloads them to other nodes to minimize task processing latency. The proposal is formulated as a joint task scheduling, channel allocation, and resource allocation problem, considering different profiles of tasks and nodes.

The IoT-edge-cloud continuum offers opportunities to support vertical industries, where many critical and time-sensitive applications require deterministic end-to-end service levels. However, most current task offloading and resource allocation proposals focus on minimizing the latency of individual tasks. This approach risks unnecessarily overloading computing and communication resources, potentially degrading overall system performance. In contrast, this paper advocates for a deterministic approach to task offloading and resource allocation where the priority is task completion before bounded deadlines (i.e., deterministic) over minimizing the latency of individual tasks. Our evaluation demonstrates that a deterministic design can effectively support a higher number of tasks through more efficient management of computing and communication resources.

### III. ARCHITECTURE AND SYSTEM MODEL

Subnetworks can be integrated in the IoT-edge-cloud continuum for local connectivity [1][11]. Subnetworks can be defined as short-range, low power radio cells, located at the edge of the cellular network. They are designed to provide highly localized and high-performance wireless connectivity for example, in vehicles, manufacturing modules or robots. Given the critical nature of some of these local connectivity scenarios, subnetworks must be able to operate standalone or connected to a parent cellular network, which can support the operation and configuration of subnetworks. Fig. 1 depicts the envisioned IoT-edge-cloud network architecture integrating subnetworks alongside communication and computing domains (edge and cloud nodes). The subnetwork consists of Subnetwork Elements (SNEs), Low Capability (LC) units, and High Capability (HC) units [11]. The HC unit has advanced networking and computational capabilities. It serves as the central hub within the subnetwork and as a gateway between the subnetwork and the parent network. The HC unit can handle most computationally intensive tasks and offer computing resources to other units within the subnetwork. An LC unit has reduced networking and computing capabilities compared to the HC unit. It can act as an aggregator or gateway between SNEs and the HC but may not have direct access to the parent network. SNEs are computationally constrained devices characterized by limited form factor, cost footprint and power/energy resources; examples include sensors and actuators. Fig. 2 illustrates an example of in-vehicle subnetworks using a zonal E/E architecture as envisioned in the transition towards software defined vehicles and autonomous driving. The in-vehicle subnetwork interconnects all devices and automotive domains with the HPCU (High-Performance Computing Unit) acting as the HC unit, 4 Zone ECUs (Zone Control Units) functioning as LC units, and sensors and actuators serving as SNEs. We assume that the subnetwork includes $N_{SNE}$ SNEs, $N_{LC}$ LC units, and $N_{HC}$ HC units, which are wirelessly interconnected following the architecture in Fig. 1. The HC unit can directly connect to the parent network, which integrates edge computing and connects to a cloud computing server. We assume $\alpha_{SNE}\%$ of tasks in the subnetwork are generated by SNEs, $\alpha_{LC}\%$ by LC units, and $\alpha_{HC}\%$ by HC unit. Tasks generated by SNEs are initially sent to LC units for processing, but may be offloaded for processing to the HC or the edge/cloud servers by the joint task offloading and resource allocation strategy. Tasks generated by an LC unit can be processed locally, or be offloaded to the HC, edge or cloud. Similarly, tasks generated by the HC can be processed locally or offloaded to the edge or cloud servers.

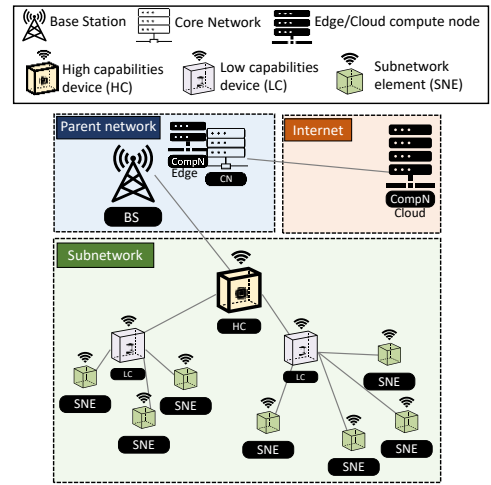

Fig. 1. Subnetworks IoT-edge-cloud network architecture.

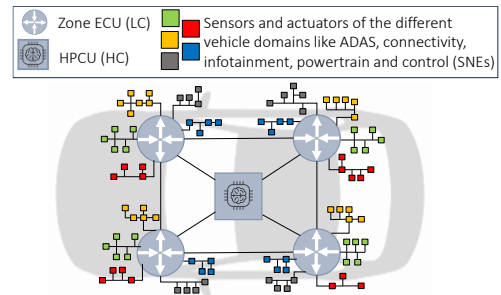

Fig. 2. In-vehicle subnetworks.

We consider a set of different tasks $f_i$, $i\epsilon\{1,2,...,I\}$ where, for the sake of generality, each task has a random generation time $t_i$. Each task is characterized by a computational demand $c_i$ and an associated size $s_i$. If a task is offloaded to a different unit from the one where it was generated, the processed result must be transmitted back to the original unit. The size of processed results $s'_i$ is smaller than the initial size of the task $s_i$. Each task has a deadline $T_i^{max}$ by which all processing must be completed. If the task is processed in a different unit from the one it was created, the deadline includes the time to transmit the task to the corresponding processing unit, the processing time, and the time to transmit the processed result back to the original unit.

Each computing unit has a processing power $P_x$ and a maximum processing capacity $C_x^{max}$ over a time period $T$ with $C_x^{max} = P_x * T$, where $x$ indexes the computing units and $x\epsilon\{Lo, E, C\}$ indicates whether the unit is local ($Lo$), edge ($E$), or cloud ($C$). The computing time for a task $f_i$ processed by computing unit $x$ is given by:

$$t_p^i = \frac{c_i}{P_x}. \qquad (1)$$

We assume an Orthogonal Frequency Division Multiple Access (OFDMA) radio access interface for the wireless links within subnetworks and for connecting subnetworks to the parent network. However, the connections within the subnetwork and those from the subnetwork to the parent network use different communication resources. We consider there are $K_1$ orthogonal communication resources for intra-subnetwork communication and $K_2$ for connecting to the parent network. In accordance with the 3GPP 5G NR standard, our model adopts a subcarrier spacing (SCS) of 30 KHz [14] and a time slot duration of 0.5 ms, which is suitable for the high-frequency communications expected in future cellular systems [1]. We denote the available data rate at any given time for communication resource $k$ in link $l\epsilon\{1,2,...,L\}$ as $r_l^{(k)}(t)$, which can be expressed as follows:

$$r_l^{(k)}(t) = \text{BW} \times \log_2(1 + \gamma_l(t))(1 - \text{BER}), \qquad (2)$$

where $BW$ is the resource's bandwidth, $\gamma_l(t)$ is the Signal-to-Interference-plus-Noise Ratio (SINR), and $BER$ is the bit error rate, which is a function of the modulation and coding scheme used. We consider various modulation schemes (BPSK, QPSK, 16-QAM, 64-QAM, and 256-QAM) and select the appropriate scheme based on the SINR level. To model channel fading, we assume a Rayleigh distribution. The data rate of a communication link $l$ is the sum of the data rates for all communication resources used in this link:

$$r_l(t) = \sum_k r_l^{(k)}(t). \qquad (3)$$

If task $f_i$ with size $s_i$ needs to be offloaded, its transmission time over a communication link $l$ is computed as:

$$t_c^i = \sum_{l\in l_i} \frac{s_i}{r_{l,i}(t)}. \qquad (4)$$

Similarly, the transmission time over a communication link $l$ for the processed result of a task can be expressed as:

$$t'^i_c = \sum_{l\in l_i} \frac{s'_i}{r_{l,i}(t)}. \qquad (5)$$

The time $T_i$ necessary to execute a task includes the processing time at the corresponding unit, and when applicable, the communication time to transmit the task from its source unit to the processing unit, and the communication time to transmit the processed result back to the source unit:

$$T_i = t_p^i + t_c^i + t'^i_c. \qquad (6)$$

A task is satisfactorily served if the total execution time $T_i$ is smaller than or equal to the task's deadline $T_i^{max}$.

## IV. RESOURCE ALLOCATION ALGORITHM

Existing task offloading and resource allocation schemes mostly focus on minimizing the total execution time of tasks. However, this approach can put excessive strain on the network, generating peaks in computing and communication demands that overload certain parts of the network. Such overloads can create bottlenecks that may unnecessarily delay the timely execution of certain tasks. In contrast, we advocate for a deterministic approach, where communication and computing resources are jointly allocated and managed to meet tasks' specific bounded latency deadlines instead of simply minimizing execution time. This approach leverages diverse tasks' deadlines to increase the number of satisfactorily executed tasks (i.e., their execution time $T_i$ is lower than their deadlines $T_i^{max}$) without overburdening the network's computing and communication resources. The proposal is designed and evaluated within the IoT-edge-cloud continuum framework described in Section III, where tasks can be processed locally or offloaded to edge or cloud servers. In this study, local processing refers to processing within a subnetwork, and tasks may be offloaded within subnetwork units, as previously described.

The objective function for our deterministic joint task offloading and resource allocation proposal is defined as:

$$min \sum_i K\left(\frac{T_i}{T_i^{max}}\right), \quad (7)$$

where $T_i$ is the execution time of task $i$, $T_i^{max}$ is the deadline of task $i$, and K is a penalty function defined as:

$$K(x) = \begin{cases} 0, & 0 \leq x \leq 1, \\ M, & x \geq 1, \end{cases} \quad (8)$$

where x = 0 represents the task's generation time $t_i$, x = 1 represents its deadline $T_i^{max}$, and $M$ is a high positive constant value. The objective is to minimize the number of tasks where the execution time $T_i$ exceeds the deadline $T_i^{max}$. The penalty function assigns a high penalty to tasks that exceed their deadlines, while tasks completed before their deadlines incur no penalty regardless of their specific execution time. The objective function aims to ensure task execution within bounded deadlines (i.e., deterministic) without placing unnecessary strain on network resources.

The objective function includes four additional constraints. First, the allocation of tasks to processing units is binary, meaning each task is assigned to a single computing unit and cannot be split across multiple units. This is expressed as:

$$\sum_{j=1}^{J} a_i^{(j)} + \sum_{q=1}^{Q} a_i^{(q)} + a_i^{(c)} = 1, \quad \forall i, \quad (9)$$

where $J$ is the number of local computing units in the subnetwork (i.e., the sum of LC and HC units) and $Q$ is the number of edge computing units. $a_i^{(j)}$, $a_i^{(q)}$ and $a_i^{(c)}$ are binary variables equal to 1 if task $i$ is allocated to the local computing unit $j$, the edge unit $q$, or the cloud unit $c$, respectively, and 0 otherwise. In line with OFDMA, the second constraint establishes that communication resources can be used by a single communication link at a time, ensuring there is no interference within the subnetwork or in the parent network. This is expressed as:

$$\sum_{i=1}^{I} b_{l,i}^{(k)} = 1, \quad \forall k, l, \quad (10)$$

where $I$ is the number of tasks, and $b_{l,i}^{(k)}$ is a binary variable equal to 1 when communication resource $k$ is allocated to transmit task $i$ in link $l$. The third constraint is that the transmission rate for all tasks utilizing one link must not exceed the maximum possible data rate of that link:

$$\sum_{i=1}^{I} r_{l,i}(t) \leq r_l(t), \quad \forall l, \quad (11)$$

where $r_{l,i}(t)$ is the data rate of link $l$ for transmitting task $i$ and $r_l(t)$ is the maximum possible data rate of link $l$. The fourth constraint is that the total processing workload of tasks assigned to a computing unit over a given time interval must not exceed the maximum processing capacity of that unit:

$$\sum_{i=1}^{I} c_i a_i^{(x)} \leq C_x^{max}, \quad \forall x. \quad (12)$$

The optimization problem formulated in equation (7) is NP-complete, and its computational complexity increases exponentially as the number of computation and communication resources grows. We have implemented a genetic algorithm in MATLAB to resolve the resource allocation problem that uses 10 generations/iterations. The implemented genetic algorithm considers a population size of 1000 for each generation and introduces mutation to converge to an (near) optimal solution [12].

## V. PERFORMANCE EVALUATION

### A. Evaluation Scenario

The IoT-edge-cloud continuum scenario considered for evaluation includes a subnetwork with 5 local processors (1 HC processor and 4 LC processors) and 35 SNEs, 1 edge node and 1 cloud server. Based on features of commercial off-the-shelf products [13], the processing power of the units are: 2.5 GHz for LC, 5 GHz for HC, 70 GHz for the edge node, and 150 GHz for the cloud server. This study does not focus on a specific IoT application. Instead, we consider that $\alpha_{SNE} = 60\%$, $\alpha_{LC} = 20\%$, and $\alpha_{HC} = 20\%$ of the tasks are generated by the SNE, LC, and HC, respectively. The tasks are generated randomly, following a uniform distribution, throughout the simulation time. Tasks can be allocated to any unit across the continuum. The processing workload and size of the tasks are modeled as uniform random variables [4] within the ranges (20, 50) M cycles [15] for the workload and (0.75, 2.25) M bits for the size [5]. The size of the processed results for a task is set to 15% of the task size. Tasks are randomly assigned a deadline $T_i^{max}$ following a uniform distribution between 20 ms and 100 ms. Without loss of generality, the penalty value M in (8) assigned to tasks that are not completed by their deadlines is 100. The bandwidths for intra-subnetwork links and for connecting to the parent network are 100 MHz and 50 MHz, respectively [14]. The intra-subnetwork links operate with a SINR of 30 dB, while the SINR for the connection between the subnetwork and the parent network varies between 3 dB and 27 dB.

We compare the performance of our proposed deterministic scheme against a random allocation and a state-of-the-art benchmark scheme. However, most of the comparisons focus on the deterministic and benchmark scheme as they outperform the random allocation. The random allocation scheme selects the computing unit for each task randomly and, if applicable, also randomly selects the communication resources of the link to reach that computing unit from those available until the task's deadline. The benchmark scheme ([4]) allocates tasks to computing units with the objective of minimizing the execution time of individual tasks based on the following objective function:

$$min \sum_i T_i, \quad (13)$$

where $T_i$ is the execution time of task $i$ as defined in eq. (6). The deterministic and benchmark schemes are compared for the same number of generations, ensuring a fair comparison by maintaining identical run times.

### B. Results

Fig. 3 compares the average ratio of satisfied tasks as a function of the number of tasks being executed. This ratio is defined as the proportion of tasks successfully completed before their deadlines relative to the total number of tasks. Results are shown for good SINR conditions in the connection to the parent network (27 dB) and bad conditions (3 dB). Fig. 3 demonstrates that our deterministic proposal achieves the highest ratio of satisfied tasks, regardless of the number of tasks or the channel quality conditions. As expected, the ratio of satisfied tasks decreases for both schemes as the number of tasks increases, due to limitations in available computing and communication resources. However, Fig. 3 clearly shows that prioritizing task completion before deadlines (i.e., 'Deterministic'), rather than trying to minimize the execution time of individual tasks (i.e., 'Minimum'), enables the system to satisfactorily handle more tasks. The same trend is observed in Fig. 4, which plots the average ratio of satisfied tasks as a function of the average SINR for the connection between the subnetwork and the parent network. The average SINR of the intra-subnetwork wireless links is maintained at 30 dB. Results are reported for scenarios with 35 and 45 tasks, as these workloads approach the capacity limit of the system model, based on Fig. 5. Fig. 5 shows the number of satisfied tasks as a function of the SINR when the total number of tasks executed is 75. The figure reveals that the number of satisfied tasks does not exceed 52, even under good channel conditions. Fig. 4 shows that the ratio of satisfied tasks increases as the link quality improves. This is because the data rate of the link increases with better SINR conditions due to the use of higher-order modulation and coding schemes. As the data rate increases, more tasks can be offloaded to the edge or cloud and be served within their latency limits. Fig. 3 and Fig. 4 show that the gains of the deterministic proposal over the benchmark scheme are more pronounced when communication or computing resources are more constrained, for instance, when the SINR to the parent network degrades or when the system approaches its capacity limit (i.e. between 45 and 55 tasks).

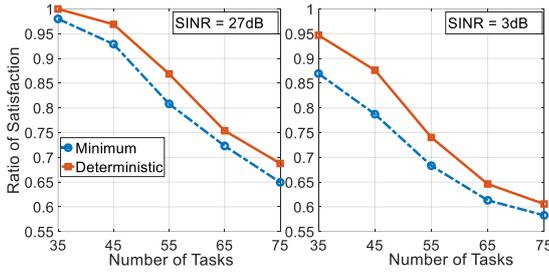

Fig. 3. Average ratio of satisfied tasks as a function of the number of tasks for SINR= 27dB and 3dB.

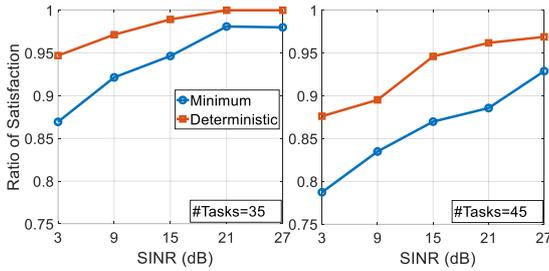

Fig. 4. Average ratio of satisfied tasks as a function of the SINR when 35 and 45 tasks are executed.

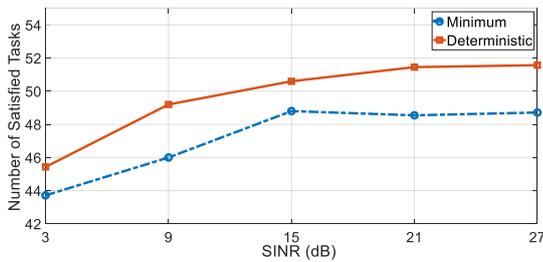

Fig. 5. Number of satisfied tasks as a function of the SINR. The total number of executed tasks is 75.

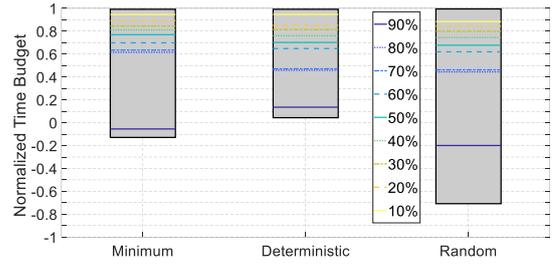

Fig. 6. Normalized time budget for 35 tasks and SINR =27 dB.

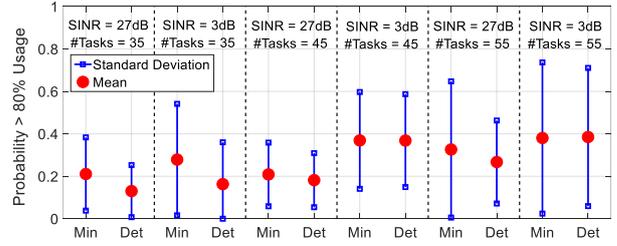

Fig. 7. Average and standard deviation of the probability of saturating the use of communication resources for various SINR and tasks.

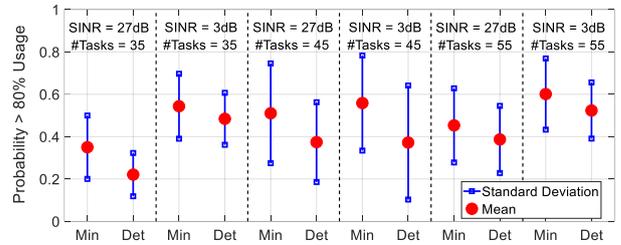

Fig. 8. Average and standard deviation of the probability of saturating the use of computing resources for various SINR and number of tasks.

Fig. 6 depicts the normalized time budget, defined as the time remaining since a task is completed to its deadline, relative to the deadline. A normalized time budget of 0 indicates that a task has been successfully completed just at its deadline, while higher values indicate that the task was completed earlier than the deadline. This metric assesses which scheme prioritizes early task completion, and which one achieves more balanced completion times. The figure presents the results as a heat bar, showing the normalized time budget for different percentiles of tasks. Fig. 6 reveals that the benchmark scheme, which aims to minimize the execution time of tasks, increases the number of tasks completed earlier compared to the deterministic scheme. For example, with the benchmark scheme, 80% of tasks are successfully completed before the normalized time budget reaches 0.6. In contrast, the deterministic proposal completes approximately 60% of tasks before the normalized time budget reaches 0.6. However, by flexibly exploiting varied task deadlines, the deterministic proposal can complete all tasks before their respective deadlines. In comparison, the benchmark scheme, focused on minimizing individual task execution times, can only successfully complete less than 90% of tasks under the conditions reported in Fig. 6. Furthermore, the results show that the benchmark scheme and the random allocation require 12% and 72% more time, respectively, to complete all tasks compared to the deterministic proposal.

Fig. 7 and Fig. 8 show the probability of high saturation in the usage of communication and computing resources, respectively. This probability represents the likelihood that communication or computing resources are utilized beyond 80% of their total capacity when new tasks arrive. Results are presented for different combinations of the number of tasks and SINR values. The figures demonstrate that the deterministic proposal (Det in figures) manages communication and computing resources more effectively compared to the benchmark scheme (Min in figures), as it reduces the probability of resource saturation. This is particularly significant because higher saturation probabilities increase the risk that new tasks will lack the necessary communication or computing resources to meet their deadlines. The figures show that the risk of saturation decreases as the SINR improves or the task load decreases. It should also be highlighted that the standard deviation of the saturation probabilities is smaller under better SINR conditions. The highest variability is observed when the number of tasks is 45 or 55, as these workloads are close to the capacity limit (Fig. 5).

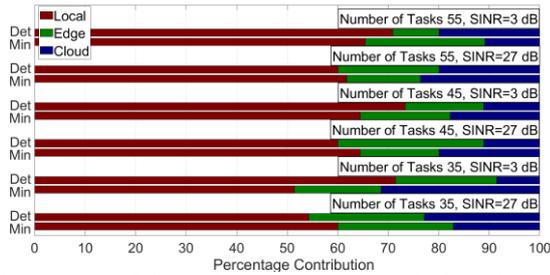

Fig. 9. Percentage of allocated tasks to computing units for various SINR values and number of tasks.

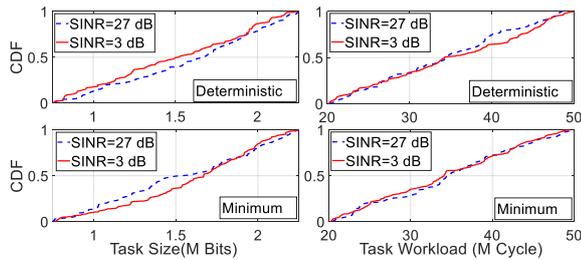

Fig. 10. CDF of size (left) and workload (right) of offloaded tasks to 6G network in different link qualities (SINR=3dB and SINR=27dB).

Fig. 9 shows the percentage of tasks allocated to different computing units in the local/IoT-edge-cloud continuum under both good and bad SINR conditions for the connection to the 6G parent network, and for different task loads. The figure shows that when SINR = 3 dB, the deterministic proposal offloads fewer tasks to the edge and cloud compared to the benchmark scheme. This is because poor link quality conditions increase the risk of saturating communication resources, as more robust modulation and coding schemes required in such conditions reduce the achievable link data rates. In contrast, the benchmark scheme attempts to offload more tasks to the edge and cloud due to their higher processing power, which ultimately penalizes the system's ability to satisfactorily support more tasks (Fig. 3 and Fig. 4). On the other hand, when SINR = 27 dB, the deterministic scheme offloads more tasks to the edge and cloud compared to the benchmark scheme and to the scenario with the lower SINR, demonstrating its capacity to adapt its offloading and allocation decisions based on the operating conditions. Fig. 10 compares the distribution of task size and workload offloaded to the parent network when the number of tasks is 45, which is close to the system's capacity. Fig. 10 (left) shows that the deterministic scheme offloads smaller tasks when the quality of the link to the network is low (SINR = 3dB), as it adapts to the reduced data rate caused by the low SINR. In contrast, the benchmark scheme offloads larger tasks under the same conditions. Fig. 10 (right) shows that the deterministic scheme offloads tasks with higher workloads to the 6G network when the link quality is poor, whereas the benchmark scheme shows no such preference. These results show that the deterministic scheme intelligently offloads tasks with higher workloads and smaller sizes under low SINR conditions. Consequently, the few tasks that can be offloaded to the network when the link data rate is low are those that will benefit the most from the higher processing power of the edge and cloud, maximizing resource utilization.

## VI. Conclusions

This study has presented a deterministic task offloading and resource allocation scheme for the joint management of communication and computing resources in the IoT-edge-cloud continuum that integrates subnetworks for local communications. The proposed scheme prioritizes meeting task deadlines over minimizing the latency in the execution of individual tasks. The scheme leverages variable latencies across tasks to satisfactorily support a higher number of tasks through a more efficient and balanced management of computing and communication resources. The proposed scheme also prevents high saturation in the usage of resources, avoiding bottlenecks in the system. This balanced management makes the system more resilient across varying computing workloads and communication quality conditions. Further analysis of the scalability of deterministic task offloading and resource allocation schemes is planned for future work.